\documentclass[12pt]{iopart}
\usepackage{iopams}  
\usepackage{amssymb,color}
\usepackage{graphicx}
\usepackage{graphics}
\usepackage{dcolumn}
\usepackage{bm}
\usepackage[normalem]{ulem}
\usepackage{cite}
\usepackage{citesort}

\def\DEGC{$^\circ$C}
\def\OUT{$(6\sqrt3\times6\sqrt3)R30^{\circ}$}
\def\6R3{$6\sqrt3$}
\def\R3{$(\sqrt3\times\sqrt3)R30^{\circ}$}
\def\3BY3{$(3\times3)$}

\def\SIG{$\sigma$}
\def\PI{$\pi$}

\def\KAY{cm$^{-1}$}

\begin{document}

\title[Contribution of the buffer layer to the Raman spectrum ...]{Contribution of the buffer layer to the Raman spectrum of epitaxial graphene on SiC(0001)}

\author{F~Fromm$^{1}$, M~H~Oliveira~Jr.$^{2}$, A~Molina-S\'anchez$^{3,4}$, M~Hundhausen$^{1}$, J~M~J~Lopes$^{2}$, H~Riechert$^{2}$, L~Wirtz$^{3,4}$ and T~Seyller$^{5}$} 
\address{$^1$~Lehrstuhl f\"ur Technische Physik, Universit\"at Erlangen-N\"urnberg, Erwin-Rommel-Str. 1, 91058 Erlangen, Germany}
\address{$^2$~Paul-Drude-Institut f\"ur Festk\"orperelektronik, Hausvogteiplatz 5-7, 10117 Berlin, Germany}
\address{$^3$~Physics and Material Sciences Research Unit, University of Luxembourg, Campus Limpertsberg, L-1511 Luxembourg}
\address{$^4$~Institute for Electronics, Microelectronics, and Nanotechnology (IEMN), CNRS UMR 8520, Dept. ISEN, 59652 Villeneuve d'Ascq Cedex, France}
\address{$^5$~Institut f\"ur Physik, Technische Universit\"at Chemnitz, Reichenhainer Str. 70, 09126 Chemnitz, Germany}
\ead{thomas.seyller@physik.tu-chemnitz.de}

\begin{abstract}
We report a Raman study of the so-called buffer layer with {\OUT} periodicity which forms the intrinsic interface structure between epitaxial graphene and SiC(0001). We show that this interface structure leads to a non-vanishing signal in the Raman spectrum at frequencies in the range of the D- and G-band of graphene and discuss its shape and intensity. \textit{Ab-initio} phonon calculations reveal that these features can be attributed to the vibrational density of states of the buffer-layer.\\
\end{abstract}


\vspace{2pc}



\section{Introduction}

Raman spectroscopy is a very powerful tool for investigating carbon materials and is intensively used for the characterization of graphene obtained by diffent methods.\cite{jorio2011a} For example, Raman spectroscopy has been shown to be extremely useful in order to discern monolayer graphene from bilayers and multilayers.\cite{ferrari2006a,graf2007a} Furthermore, this technique provides information about the carrier concentration in graphene,\cite{das2008a} the edges of graphene flakes,\cite{pimenta2007a,gupta2009a,zhang2011a} and about the properties of graphene nano ribbons.\cite{gillen2010a,gillen2010b,ryu2011a} Hence it is no surprise that it is also used to investigate epitaxial graphene grown on silicon carbide. \cite{faugeras2008a,roehrl2008a,ni2008a,ferralis2008a,lee2008a,robinson2009a,robinson2009b,domke2009a,speck2011a,oliveira2011a,emtsev2009a,tiberj2011a,ferralis2011a}

\begin{figure}[t]
\centering
\includegraphics[width=12cm]{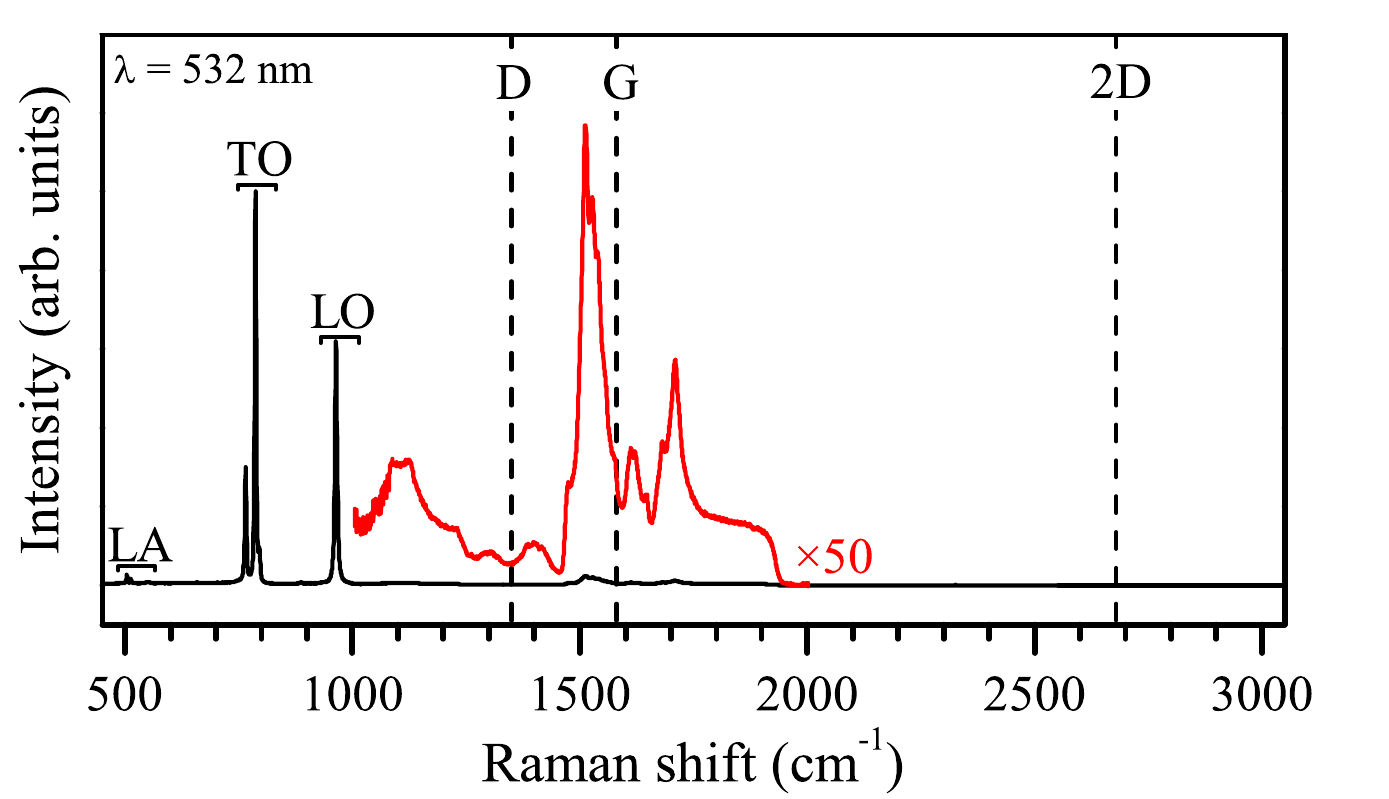}
\caption{\label{fig:raman} Raman spectrum of 6H-SiC measured with a laser wavelength of 532~nm. The signals due to the longitudinal acoustic (LA), transverse optical (TO), and longitudinal optical (LO) phonon modes are indicated. The region between 1000 and 2000~{\KAY} is also shown after multiplication by a factor of 50. In this region, the signals due to two-phonon processes are visible. Three-phonon processes are very weak and can be neglected. The positions of the D, G, and 2D line of graphene are indicated by the vertical dashed lines.}
\end{figure}

The Raman spectrum of graphene usually shows three main features: the D-band at around 1350~{\KAY}, the G-band at about 1580~{\KAY}, and the 2D-band at approximately 2680~{\KAY}. However, Raman spectroscopy is not a surface sensitive method and usually the probed sample volume is much larger, i.e. deeper, than the graphene sheet itself. This leads to the presence of substrate-related features in the spectrum as well. For many substrates such as SiO$_2$/Si this is not a problem because these features do not overlap with the graphene signals. This is different for epitaxial graphene grown on SiC where the Raman spectrum in the D- and G-range is dominated by the two-phonon modes of the SiC substrate.\cite{roehrl2008a} As an example we show in figure \ref{fig:raman} a Raman spectrum of 6H-SiC. The region between around 1000 and 2000~{\KAY} is dominated by two-phonon processes. Therefore, it has become common to correct the spectra of epitaxial graphene on SiC by subtracting the spectrum of the bare substrate. This procedure, however, assumes that the spectrum contains only contributions from the epitaxial graphene and from the SiC bulk. In the case of epitaxial graphene on SiC(0001) this assumption, however, may not be correct because it is  known that the graphene sheet resides on the so-called buffer layer\cite{emtsev2007a,emtsev2008a}. It is nowadays widely accepted that the buffer layer itself is a graphene-like honeycomb lattice of carbon atoms ontop of an otherwise unreconstructed Si(0001) surface. \cite{emtsev2007a,emtsev2008a} The buffer layer shows the undistorted \SIG-states of graphene but a distorted \PI-band. The distortions are caused by the hybridization of the $\pi$-states with the states of the SiC(0001) surface and by the formation of covalent bonds between some of the graphene-carbon atoms and underlying silicon atoms. Several theoretical studies have investigated various aspects of this structure, all in good agreement with experimental results.\cite{varchon2007a,mattausch2007a,mattausch2007b,mattausch2008a,kim2008a}  It is natural to ask about the contribution of the buffer layer to the total Raman spectrum measured from epitaxial graphene on SiC. In this paper we will show that the buffer layer leads to a non-negligible contribution in the Raman spectrum, and we will discuss the origin of the signal.

\section{Experimental details}

\begin{figure}[t]
\centering
\includegraphics[width=12cm]{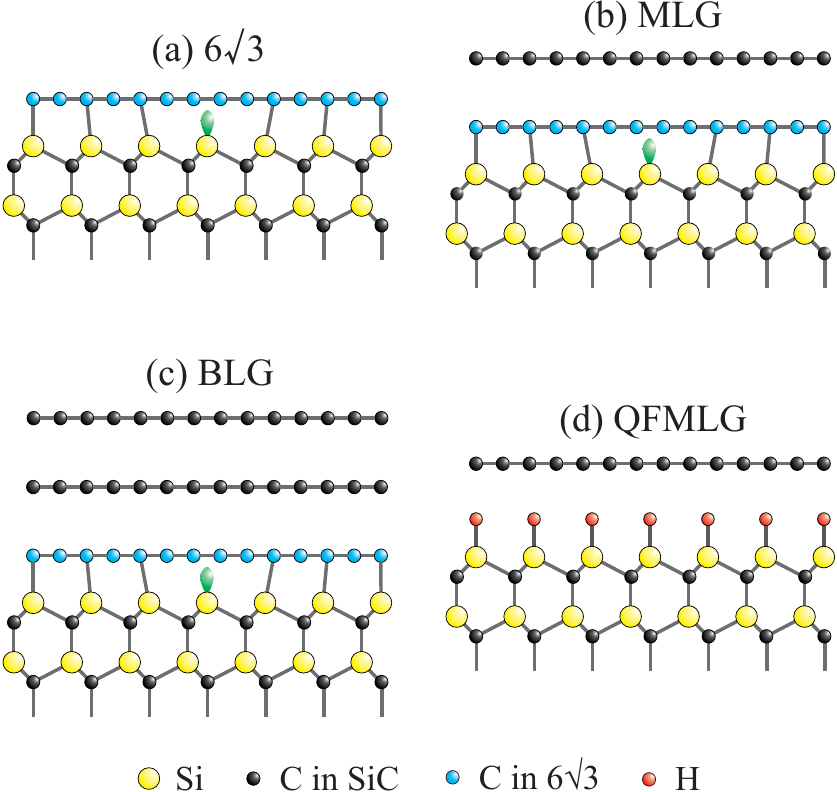}
\caption{\label{fig:structures} Schematic structures (side view) of the samples used in the present study. (a) The buffer layer (\6R3) with \OUT periodicity. (b) Monolayer graphene (MLG) situated on the buffer layer. (c) Bilayer graphene (BLG) situated on top of the buffer layer. (c) Quasi-free standing monolayer graphene (QFMLG) on top of the hydrogen-saturated SiC surface. The carbon atoms of the buffer layer are plotted in blue.}
\end{figure}

In order to identify the contribution of the buffer layer to the Raman signal of epitaxial graphene we have studied different samples. The structures of the samples are depicted schematically in fig. \ref{fig:structures}. All samples were prepared on chips cut from nitrogen-doped, on-axis oriented 6H-SiC(0001) wafer purchased from SiCrystal AG. Despite the fact that the wafer had an epi-ready chemo-mechanical polish (CMP), the surfaces were treated with an additional hydrogen etch in 1 bar H$_2$ at 1500~{\DEGC}.\cite{ostler2010a} Samples covered with the buffer layer (\6R3~for short, see fig. \ref{fig:structures}(a)) were prepared by annealing the SiC(0001) sample in 1 bar Ar at $T=1450$~{\DEGC}.\cite{ostler2010a} Monolayer graphene on the buffer layer (termed MLG, see fig. \ref{fig:structures}(b)) was obtained by annealing the SiC substrate in 1 bar Ar at 1650~{\DEGC}. \cite{ostler2010a,emtsev2009a} From previous studies it is known that such samples may contain inclusions of bilayer graphene at positions close to the step edges.\cite{emtsev2009a} Therefore, our micro-Raman spectroscopy measurements (see below) also allowed us to obtain Raman spectra from bilayer graphene on the buffer layer (BLG for short, see fig. \ref{fig:structures}(c)). Finally, quasi-free standing graphene on hydrogen-saturated SiC(0001)\cite{riedl2010b,speck2010a,speck2011a,forti2011a} (QFMLG, see fig. \ref{fig:structures}(d)) was obtained by annealing samples covered by the buffer layer in 1 bar hydrogen.\cite{speck2011a} Reference spectra of 6H-SiC were obtained from a hydrogen etched sample which is free of any carbonaceous surface layer.

The samples prepared in the above mentioned ways were thorougly characterized by X-ray photoelectron spectroscopy (XPS) and atomic force microscopy. Micro-Raman spectroscopy measurements were performed using a Jobin Yvon T64000 triple spectrometer combined with a liquid nitrogen cooled CCD detector. A frequency doubled Nd:YVO$_4$ laser with a wavelength of 532~nm was employed. Additional spectra were measured using an Ar ion laser providing wavelengths of 476~nm and 514~nm. The laser beam was focused onto the sample by a 100x objective with numerical aperture $\mbox{NA}=0.9$ and the scattered light was detected in backscattering geometry. The laser spot size was 1~$\mathrm{~\mu m}$. Unless otherwise stated, the Raman spectra were measured under the same conditions. The raw data was normalized to the maximum of the TO phonon mode of 6H-SiC at about 780$\;$cm$^{-1}$. 

\section{Results and Discussion}

\begin{figure}[t]
\centering
\includegraphics[width=12cm]{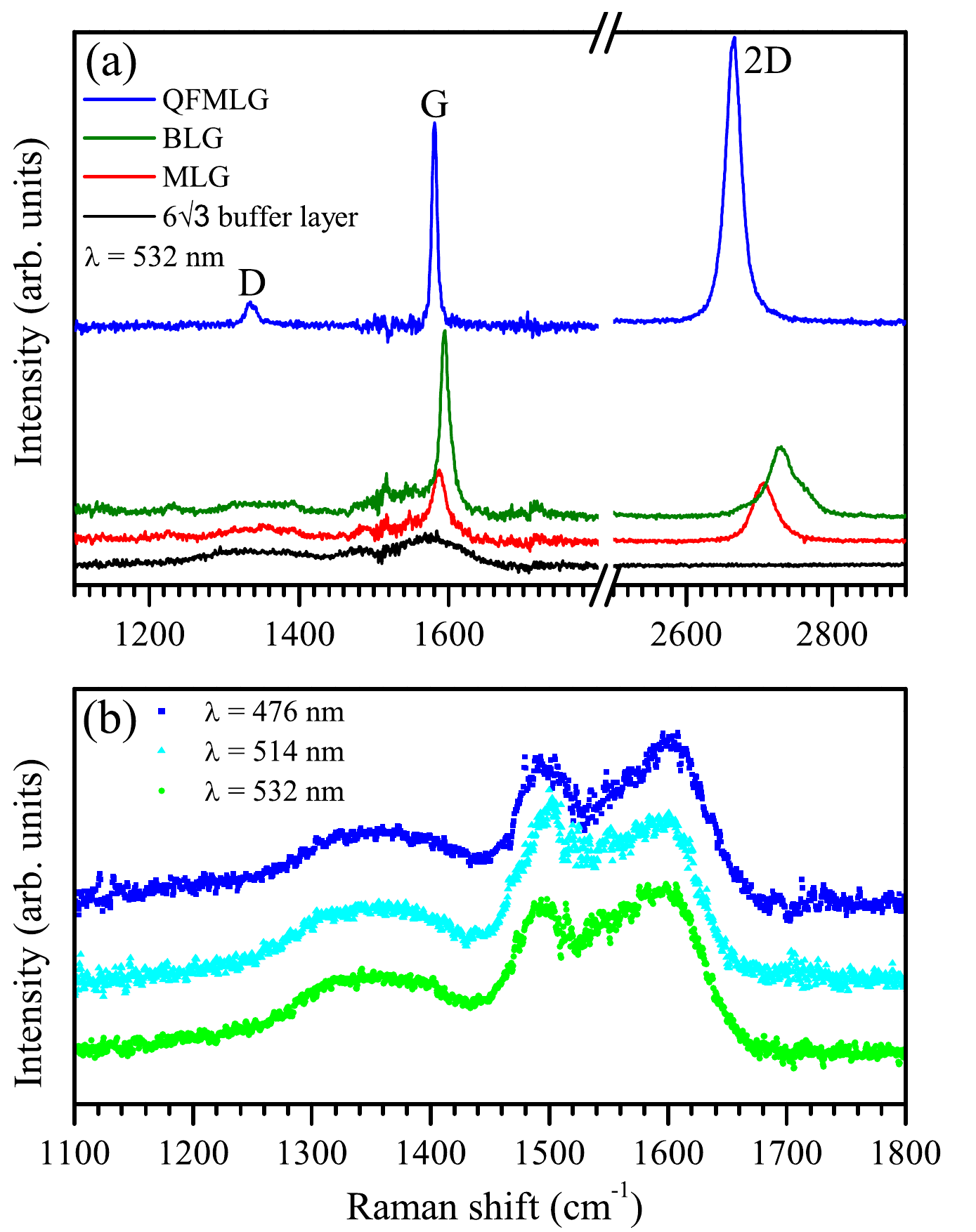}
\caption{\label{fig:spectra} (a) Raman spectra of (from bottom to top) the buffer layer (\6R3), monolayer graphene on the buffer layer (MLG), bilayer graphene on the buffer layer (BLG), and quasi-free standing monolayer graphene (QFMLG) on H-terminated SiC(0001). The spectra were measured with a laser wavelength of 532~nm. (b) Raman spectra of the buffer layer measured with three different laser energies.}
\end{figure}

Figure \ref{fig:spectra}(a) compiles typical Raman spectra of the different samples described above. The spectra were collected at a laser wavelength of 532~nm. The lowest spectrum (labelled \6R3) was measured on the sample covered by the buffer layer. At low energies  that spectrum contains two rather broad features, one centered at around 1350~{\KAY} and one at 1580~{\KAY}. The latter is accompanied by a smaller peak at the low-energy side situated at around 1485-1490~{\KAY}.  No 2D line is observed for the buffer layer. 

The spectra of MLG and BLG are also displayed in figure \ref{fig:spectra}(a). Both spectra exhibit a G line and a 2D line.  The 2D line of MLG at 2706~{\KAY} is very well described by a single Lorentzian with a full width of half maximum of 35~{\KAY}, wich agrees with the notion that the sample is covered mainly by monolayer epitaxial graphene. The shape of the 2D band of BLG in fig. \ref{fig:spectra}(a) is consistent with what has been observed previously on both exfolitated graphene\cite{ferrari2006a,graf2007a} and epitaxial graphene.\cite{roehrl2008a} Note, that in this work we are not interested in the exact positions of the G and 2D bands, which might be influenced by strain and charge. For the discussion of this topic we refer the reader to previous published work. \cite{roehrl2008a,ni2008a,ferralis2008a,lee2008a,robinson2009a,robinson2009b,speck2011a} However, what is important for the present work is the observation that the spectra of MLG and BLG contain the same broad features between 1200 and 1665~{\KAY} that are observed for the \6R3 sample. Finally, figure \ref{fig:spectra}(a) also shows the spectrum of a sample of QFMLG, i.e. a layer of graphene on SiC(0001) without the buffer layer at the interface. The spectrum consists of three narrow lines: the D line, the G line, and the 2D line, as discussed in previous work.\cite{speck2011a} In contrast to the spectra of MLG and BLG, the broad features between 1200 and 1665~{\KAY} are absent and the spectrum is basically flat between D and G line.

This fact provides important input. As mentioned above, the spectra shown are difference spectra where the spectrum of a clean SiC sample is subtracted from that of the graphene covered one. One could therefore think that the broad features described above are the result of an insufficient background correction. This is clearly ruled out by the fact that the spectrum of QFMLG, which was obtained in exactly the same way as those of the {\6R3}, MLG, and BLG samples, does not show this features. The only effect of the background subtraction is the increase of noise on both sides of the G line which can be seen in all spectra. This can be understood by considering that at those frequencies the intensity in both data sets, the one of the sample with graphene and the one used for background subtraction, is particularily large due to the contribution of the SiC substrate (see fig. \ref{fig:raman}). The larger intensity at these frequencies leads to a larger statistical noise ($\sqrt{n}$ with $n$ being the count rate) which  of course is not removed by the subtraction of the spectra. Therefore we can savely state that the spectrum labelled {\6R3} in fig. \ref{fig:spectra} is the true Raman spectrum of the buffer layer which exists at the interface between SiC(0001) and epitaxial graphene.

Figure \ref{fig:spectra}(b) shows three Raman spectra of the buffer layer measured with three different laser wavelengths of 476~nm, 514~nm, and 532~nm, which correspond to excitation energies of 2.33~eV, 2.41 eV, and 2.60~eV, respectively. Since the scattering intensity is zero for wavenumbers larger than approximately 1665~{\KAY}, we show only the low-energy part of the spectrum between 1100 and 1800~{\KAY}. As can be seen from figure \ref{fig:spectra}(b), the observed Raman spectrum is virtually independent of the laser wavelength, i.e. the broad peaks at 1350~{\KAY}, 1485-1490~{\KAY}, and 1580~{\KAY} show no dispersion and there is hardly any change in the shape of the signals. 

\begin{figure}[t]
\centering
\includegraphics[width=12cm]{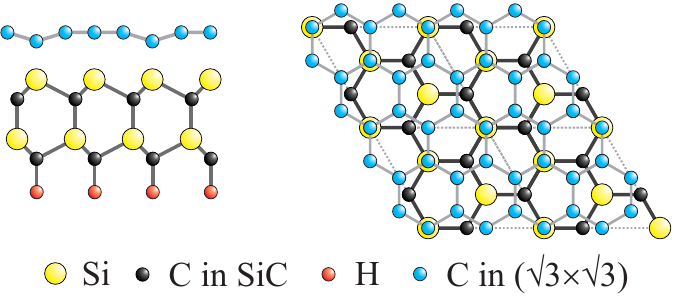}
\caption{\label{fig:model} Unit cell used in the simulation of the  buffer layer. The carbon atoms of the buffer layer are plotted in blue. For clarity, the top view shows only atoms of the topmost SiC bilayer. Hydrogen atoms passivate the dangling bonds of the carbon  atoms of SiC at the bottom. In this configuration two carbon atoms of buffer layer are on the top of silicon atoms. Drawing not to scale.}
\end{figure}

The Raman spectrum shown in Fig. \ref{fig:spectra}(b) does not seem to be composed of discrete peaks but rather resembles a vibrational density of states (vDOS). This is plausible since the unit-cell of the {\OUT} reconstruction is quite large and the corresponding reciprocal unit cell is small. Thus, a large part of the phonon-dispersion is folded back onto $\Gamma$ and becomes potentially Raman active. To a good approximation, one may therefore assume that the Raman spectrum of Fig. \ref{fig:spectra}(b) corresponds to the vDOS of the buffer layer. We have verified this hypothesis by an explicit calculation of the phonon dispersion and vDOS of the buffer layer using \textit{ab-initio} methods as presented in the following.

Since the unit-cell of the {\OUT} reconstruction (and even of the recently proposed $(5\times5)$ superstructure\cite{pankratov2010a}) is prohibitively large for \textit{ab-initio} calculations of phonons, we have chosen to work with the {\R3} reconstruction which was also used in the electronic-structure calculations of Refs. \cite{varchon2007a} and \cite{mattausch2007a}. This unit-cell corresponds to a $(2\times2)$ supercell of graphene. The unit cell for our simulation of the buffer layer on SiC is shown in Fig. \ref{fig:model}. With respect to free-standing graphene, it corresponds to a $(2 \times 2)$ unit cell, containing 8 carbon atoms.  With the aim of obtaining reliable results for the phonons of the buffer layer, the commensurability between the buffer layer and SiC is obtained by squeezing the substrate by 8~\%, adopting the experimental lattice  constant of graphene (2.46 \AA). (This is different from the procedure  in Ref.~\cite{varchon2007a} where the lattice constant of graphene was increased by 8~\% in order to match the experimental lattice constant of SiC.) The SiC substrate is simulated by four atomic layers (2 Si layers and 2 C layers),  passivated with hydrogen atoms at the bottom. Note that in this configuration two of the eight carbon atoms of the buffer layer are on top of the silicon atoms, forming a covalent bond. The atomic positions inside the unit cell have been calculated with density functional theory (DFT)\cite{kohn1965a,parr1989a}, in the local density approximation (LDA).\footnote{We note that the use of DFT with purely (semi)local functionals is questionable for the use in layered systems where Van der Waals forces are expected to play an important role. Nevertheless, the LDA seems to work fine for the calculation of geometries and even of phonon frequencies (however, not for the binding energies) of several layered systems such as graphite\cite{kresse95,wirtz2004a}, hexagonal boron nitride\cite{kern1999a}, graphene on a Nickel(111) surface\cite{allard2010a}, and MoS$_{2}$.\cite{molina2011a} This seemingly good performance is probably due to a fortuitious  error cancellation: the small (but non-negligible) covalent part of the inter-layer binding is overestimated while the Van der Waals part of the binding energy is completely neglected. For a more precise treatment of Van der Waals forces, calculating electron correlation in the random-phase approximation, we refer to the work of Marini et al. on hBN\cite{marini2006a}, Mittendorfer et al. on graphene bound to metallic substrates\cite{mittendorfer2011a} or Kim et al. for the binding of benzene molecules on a Si surface \cite{kim2012a}.} The calculations were performed with the Quantum-Espresso code\cite{QE-2009} using ultra-soft Vanderbilt pseudopotentials, a $12 \times 12 \times 1$ k-point grid, and and energy cutoff of 35 Ry. Due to the formation of covalent bonds, the carbon atoms on top of surface Si atoms display an inward buckling by $\Delta$z = 0.39 \AA (see Fig.~\ref{fig:model}). The distance between the other carbon atoms and the surface-Si plane is $d$ = 2.02 \AA.

\begin{figure}
\centering
\includegraphics[width=12cm]{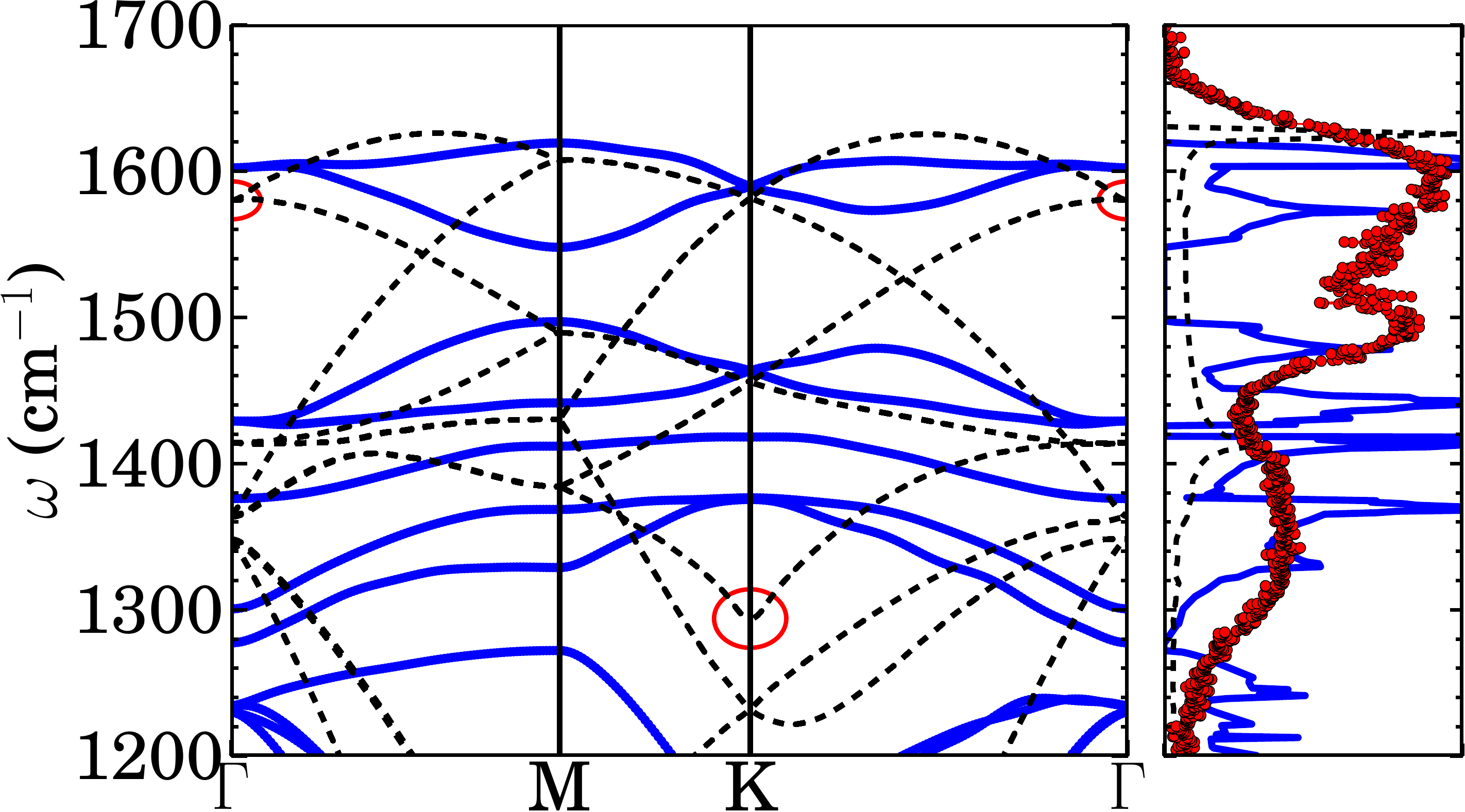}
\caption{\label{fig:phonons} Left panel: Phonon modes of the buffer layer (blue lines) and free-standing graphene (black dashed line) in a $(2 \times 2)$ unit cell. The red circles mark the Kohn anomalies at $\Gamma$ and $K$ of free-standing graphene. Right panel: vibrational density of states of the buffer layer (blue line) and of free-standing graphene (black-dashed line) in comparison with the experimental Raman data (red dots).}
\end{figure}

Starting from the optimized geometry, we use density functional perturbation theory (DFPT),\cite{gonze1997a,baroni2001a} to calculate the phonon dispersion of the buffer layer. The result is shown in Fig.~\ref{fig:phonons}, where we have concentrated on the frequency range above 1200 cm$^{-1}$ which is important for the interpretation of the spectra in Fig.~\ref{fig:spectra}(b). The broad blue lines correspond to the modes of the buffer layer (The SiC bulk modes have frequencies below 1200 cm$^{-1}$). For comparison we have included the phonon bands of isolated graphene in a $2\times2$ unit cell (containing 8 atoms, leading to 24 phonon branches marked by the black dashed 
lines).\footnote{There is the eternal question if one should use (\textit{ab-initio}) optimized lattice constants or experimental lattice constants for the phonon calculations. Since the local-density approximation tends to overbind, the optimized lattice constant is smaller than the experimental one. The calculated phonon frequencies are in general a little bit higher than the experimental values and need to be scaled down by about 1\%\cite{wirtz2004a}. We use here the experimental lattice constant of pure graphene (for both the isolated graphene and the buffer layer). In this case, the phonon frequencies are a little bit lower than the experimental ones. We thus rescale the calculated phonon dispersions (of both isolated and buffer graphene) by the respective ratio of the experimental and theoretical values of the E$_{2g2}$ (highest optical mode at $\Gamma$) phonon frequencies. For isolated graphene the Raman G-line has the value of 1580 cm$^{-1}$ (according to recent measurements on suspended graphene\cite{berciaud2009a}) and our calculated value is 1568 cm$^{-1}$. For the buffer layer on SiC(0001), the experimental value (measured by electron energy loss spectroscopy \cite{koch2013}) is 1595 cm$^{-1}$ and our calculated value is 1558 cm$^{-1}$. These differences are related to the unknown strain state of the buffer layer.} The right panel of Fig.~\ref{fig:phonons} shows the vibrational density of states (vDOS) of the buffer layer and free-standing graphene, together with the experimental Raman spectra (red dots).

The phonon dispersion of the buffer-layer is considerably changed compared to the one of isolated graphene. The changes in the electronic structure (lifting of the linear crossing at K and separation of the $\pi$ and $\pi^*$ bands by more than 2 eV \cite{varchon2007a}) lead to the elimination of the Kohn anomalies at K and $\Gamma$ (marked by red circles in Fig.~\ref{fig:phonons}), similar to the findings for graphene on a Ni(111) surface.\cite{allard2010a} A consequence of this is the absence of the 2D line (around 2680 cm$^{-1}$) in the Raman spectrum of the buffer layer, as seen in Fig. \ref{fig:spectra}. Furthermore, the buffer layer displays flatter bands than pure graphene. In particular, the overbending of the highest optical branch around $\Gamma$ almost disappears. This brings the frequency of the highest vDOS peak at
1630 cm$^{-1}$ down to 1620 cm$^{-1}$, in agreement with the highest peak in the Raman spectrum. Additionally, some degeneracies are broken at the $\Gamma$ and M points. These modifications lead to noticeable changes in the vDOS of the buffer layer. For instance, a clear gap emerges between 1500 and 1550~{\KAY}, in agreement with the minimum at the same frequency range in the Raman spectrum. The broad feature in the Raman spectrum around 1360 cm$^{-1}$ can also be associated with peaks in the 
vDOS of the buffer layer that are due to flat phonon bands. The phonon bands of pure graphene are very dispersive in this range and the corresponding vDOS is flat.

\begin{figure}
\centering
\includegraphics[width=12cm]{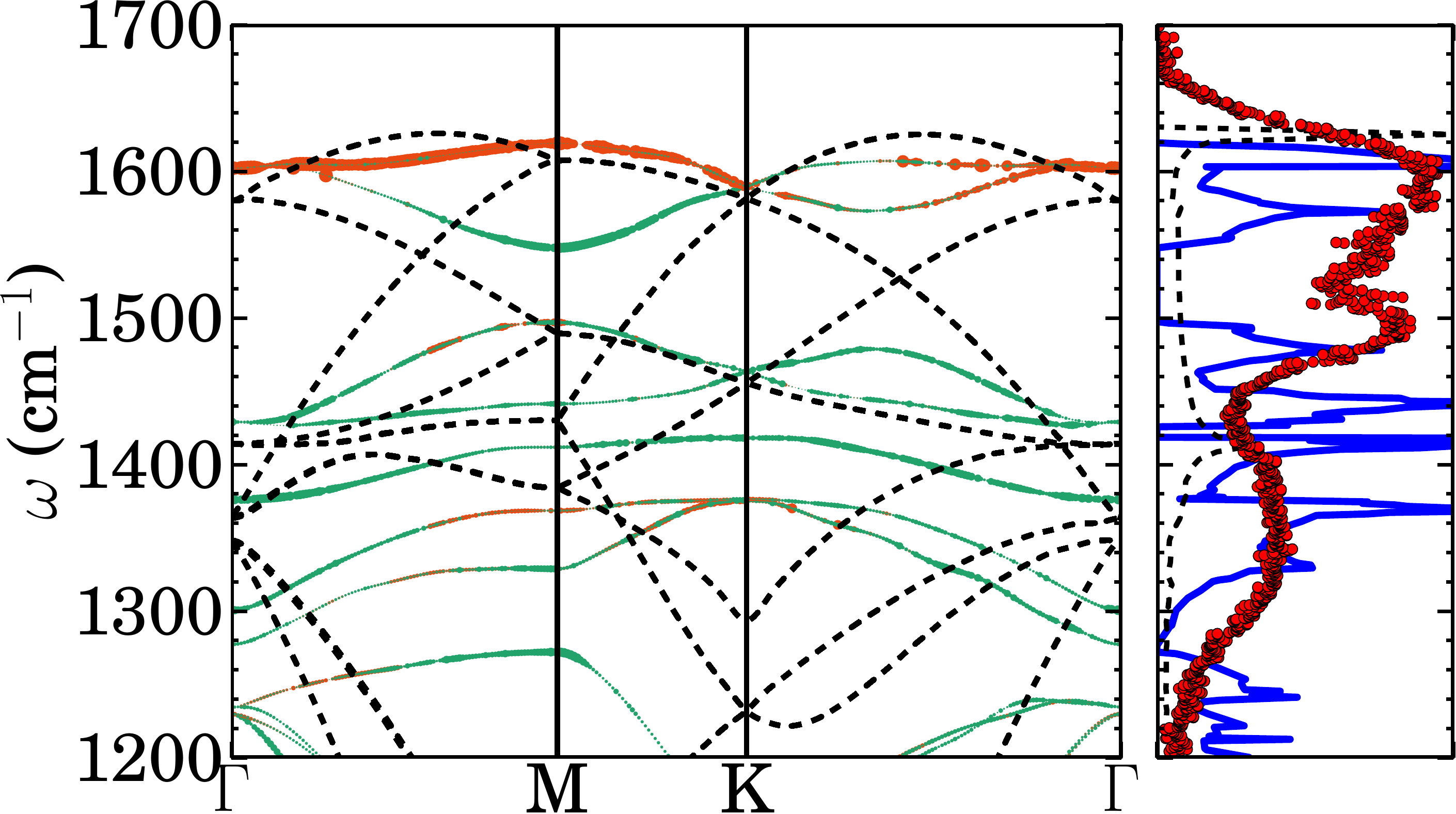}
\caption{\label{fig:projection} Left panel: Phonon bands weighted according to the contribution of two different subspaces (see main text). Right panel: same as in Fig. \ref{fig:phonons}.}
\end{figure}

For a better understanding of the gap opening between 1500 and 1550~{\KAY}, we analyze the phonon eigenvectors of the buffer layer by projecting them onto the eigenvectors of the isolated (undisturbed) graphene. In Fig. \ref{fig:projection} we have projected every eigenvector of the buffer layer onto two subspaces of eigenvectors of isolated
graphene. The color of the phonon branch indicates which subspace dominates the character of the vibration. The first subspace (orange color) is composed by the two eigenvectors of highest frequency at the respective phonon-wave vector $\bm{q}$. This definition becomes ambiguous at the crossing point of the 2nd and 3rd highest mode in between the high-symmetry points. But it is well defined at the high-symmetry points $\Gamma$, M, and K, where double-degeneracies are observed. The second subspace (green color) includes the next four eigenvectors in energy order, and it has also two 2-fold degeneracies at $\Gamma$, M, and K. We focus on the dispersion around M. We can assign the first and third phonon branches as being predominantly due to the orange subspace. The frequency difference is 100~{\KAY}. All the other eigenvectors belong to the green subspace. The perturbation of the buffer-layer vibrations by the partially covalent bonding to the SiC substrate lifts the degeneracies. The splitting is so strong that it even changes the order of the phonon modes: The lower frequency mode due to the first subspace falls below the highest frequency mode of the second subspace.
In the phonon dispersion, between $\Gamma$ and M (and also between $\Gamma$ and K), this leads to an avoided crossing between the 2nd and 3rd highest phonon mode and thus to the opening of a gap from 1500 to 1550~{\KAY}. For the lower frequency modes, similar analysis can be made. But the analysis becomes more complicated due to a large number of participating modes. In the right panel of Fig.~\ref{fig:phonons}, one can observe an approximate agreement between dips in the Raman spectrum and gaps in the vDOS. The same holds for the peaks in the Raman spectrum and in the vDOS. Since we used a simplified supercell for the buffer-layer geometry, we do not expect perfect agreement here. But we consider the present calculation as a qualitative argument that the observed features in the Raman spectra of the buffer-layer can indeed be associated with the vibrational density of states.

\section{Conclusions}

In conclusion we have been able to unambiguously identify the Raman spectrum of the buffer layer (\6R3) which exists at the interface between epitaxial graphene and SiC(0001). We have shown that it constitutes a non-negligible contribution underlying the graphene spectrum especially at frequencies around the D- and G-line. This implies that proper Raman analysis of graphene on SiC(0001) requires that the spectrum  is also corrected for the buffer layer contributions. Neglecting the buffer layer will lead to errors in the interpretation of Raman spectra of epitaxial graphene on SiC(0001). Furthermore, we have discussed the Raman spectrum of the buffer layer in terms of the vibrational density of states. To that end, \textit{ab initio} calculations on a \R3 superstructure have been performed which revealed a complete extinction of the Kohn anomally, in agreement with the lack of a Dirac cone in the electronic structure \cite{emtsev2007a,emtsev2008a} and with the absence of a 2D line in the Raman spectrum. As a consequence, phonon bands become flatter than in free-standing graphene. In addition, the carbon-silicon covalent bonds modify substantially the frequencies and lead to a mixing of the phonon branches of isolated graphene. This leads to a breaking of degeneracies in the phonon dispersion and the vDOS of the buffer layer is richer in structure than that of isolated graphene. In particular a clear gap between 1500 and 1550~{\KAY} emerges which agrees fairly well with the Raman spectrum. 

\section{Acknowledgement}

This work was supported by the European Science Foundation (ESF) in the framework of the EuroGRAPHENE project Graphic-RF (grant number 09-EuroGRAPHENE-FP-018) and by the European Union (EU) in the framework of the project ConceptGraphene (grant number 257829). A.M.-S. and L.W. acknowledge funding by the French National Research Agency (ANR) through project ANR-09-BLAN-0421-01. Calculations were done at the IDRIS supercomputing center, Orsay (Proj. No. 091827), and at the Tirant Supercomputer of the University of Valencia (group vlc44). The authors thank L. Ley, J. Ristein, and R. J. Koch for stimulating discussions.

\end{document}